# Interplay of magnetism and high-$T_c$ superconductivity at individual Ni impurity atoms in $Bi_2Sr_2CaCu_2O_{8+\delta}$

E.W. Hudson[1*], K.M. Lang[1], V. Madhavan[1], S.H. Pan[1†], H. Eisaki[2‡],

S. Uchida[2] & J.C. Davis[1]

[1]      *Department of Physics, University of California, Berkeley, CA 94720, USA.*

[2]      *Department of Superconductivity, University of Tokyo, Tokyo, Japan.*

\*       *Present Affiliation: National Institute of Standards and Technology, Gaithersburg, MD 20899, USA.*

†       *Present Affiliation: Department of Physics, Boston University, Boston, MA 02215, USA.*

‡       *Also: Department of Applied Physics, Stanford University, Stanford, CA 94205-4060, USA.*

**In conventional superconductors, magnetic interactions and magnetic impurity atoms are destructive to superconductivity[1]. By contrast, in some unconventional systems, e.g. superfluid $^3$He and superconducting $UGe_2$, superconductivity or superfluidity is actually mediated by magnetic interactions. A magnetic mechanism has also been proposed for high temperature superconductivity (HTSC) in which an electron magnetically polarizes its environment resulting in an attractive pairing-interaction for oppositely polarized spins[2-6]. Since a magnetic impurity atom would apparently not disrupt such a pairing-interaction, it has also been proposed[5,6] that the weaker influences on HTSC of magnetic Ni impurity atoms compared to those of non-magnetic Zn are evidence for a magnetic mechanism. Here we use scanning tunneling microscopy (STM) to determine directly the influence of individual Ni atoms on the electronic structure of $Bi_2Sr_2CaCu_2O_{8+\delta}$. Two local d-wave impurity-states[7,8] are observed at each Ni. Analysis of their energies surprisingly reveals that the primary quasiparticle scattering effects of Ni atoms are due to *non-magnetic* interactions. Nonetheless, we also demonstrate that a magnetic moment coexists with unimpaired superconductivity at each Ni site. We discuss the implications of these phenomena, and those at Zn[9], for the pairing-mechanism.**

In our studies we use two different $Bi_2Sr_2Ca(Cu_{1-x}Ni_x)_2O_{8+\delta}$ single crystals, grown by the floating-zone technique. These crystals have x = 0.005 with $T_c$ = 83 K and x = 0.002 with $T_c$ = 85 K, respectively. The Ni atoms substitute for Cu atoms in the superconducting $CuO_2$ plane and are believed to be in the $Ni^{2+}$ $3d^8$ electronic state, as compared to the $Cu^{2+}$ $3d^9$ of Cu. Above the superconducting transition temperature $T_c$ each Ni atom possess a strong magnetic moment of approximately 1.5 $\mu_B$[10]. The samples are cleaved in cryogenic ultra-high vacuum at 4.2 K exposing the BiO crystal plane and, when inserted into the STM head are imaged with atomic resolution.

For d-wave superconductors, theory predicts that quasi-particle scattering at an impurity atom creates a local electronic state (impurity-state) nearby[7,8,11-17]. Such states, which have been analyzed for both potential (non-magnetic)[7,8,11,12,14-16] and magnetic[8,12-17] scattering, can be thought of as almost localized quantum orbitals of well defined energy $\Omega$ and spatial structure[9, 18].

The spatial distribution of impurity-states at Ni impurity atoms can be imaged by measuring, as a function of position, the differential tunneling conductance G = dI/dV. This is proportional to the local-density-of-states (LDOS) and we refer to this process as LDOS-mapping. As an example, Fig. 1 shows two simultaneously acquired LDOS maps taken at sample bias V = ± 9 mV. They reveal both the particle-like (positive-bias) and hole-like (negative-bias) components of one of the impurity-states that exist at each Ni. At +9 mV "+-shaped" regions of higher LDOS are observed, while at the same locations but at -9 mV the corresponding higher LDOS regions are "X-shaped". LDOS maps at V = ± 19 mV show the particle-like and hole-like components of a second impurity-state at Ni whose spatial structure is very similar to that at V = ± 9 mV.

High-resolution ± 9 mV LDOS maps (Figs. 2a and 2b) acquired simultaneously with a topographic image of the BiO surface (Fig. 2c) show in detail how an impurity-state consists of two spatially complementary components. By this we mean that the particle-like LDOS is high in regions where the hole-like LDOS is low, and vice versa. To illustrate these relationships, Fig. 2d shows a schematic diagram of the relative locations of Cu atoms, the orientation of the $d_{x^2-y^2}$ superconducting order parameter (OP), and the location of the particle-like (green) and hole-like (purple) components of the impurity-state.

We also probe the energy dependence of the LDOS at several locations near a Ni atom. Figure 3 shows typical dI/dV spectra taken at locations above the Ni atom, above the first nearest neighbour (1-NN) and second nearest neighbor (2-NN) Cu atoms, at a distance of 30 Å from the impurity site, and the average spectrum for the whole impurity-state region. Two clear particle-like LDOS peaks are observed at the Ni site. The average magnitudes of these on-site impurity-state energies measured at 8 different Ni sites are $\Omega_1 = 9.2 \pm 1.1$ meV and $\Omega_2 = 18.6 \pm 0.7$ meV. As shown, both these peaks become hole-like at all the 1-NN Cu sites and again particle-like at the 2-NN Cu sites. Despite these complexities, the spatially averaged spectrum for this whole region (shown in Fig. 3b) remains very close to particle-hole symmetric.

Several important conclusions can be drawn from these data. First, the two on-site LDOS peaks reveal that there are two distinct impurity-states associated with each Ni atom. This can be explained by theories that consider both potential and magnetic interactions[8]. In a d-wave superconductor potential scattering generates a single spin-degenerate impurity-state[7,8,11]. A weak additional magnetic interaction between an impurity moment and the quasi-particle spin lifts the spin-degeneracy, creating two spin-polarized impurity-states at each magnetic impurity atom[8]. The good qualitative agreement between the data and this model indicates that Ni atoms do possess a magnetic moment in the superconducting state.

Secondly, as shown in Fig. 2 and 3, the spectral weight of an impurity-state oscillates between particle-like and hole-like as a function of distance from the Ni atom. Nevertheless, the average conductance spectrum over the whole impurity-state remains almost particle-hole symmetric. Theoretically, such overall particle-hole symmetry in an impurity-state spectrum is expected when superconductivity is not disrupted[12,14-16,19,20]. Furthermore, calculations for the d-wave impurity-state wavefunction show that the particle-like and hole-like LDOS should be spatially complementary[8,12,14-16]. The agreement of our observations with theory shows that superconductivity is not disrupted locally by the magnetic moment of Ni.

A third important observation is that the superconducting gap magnitude (as deduced from the energy of the coherence peaks) does not change as we approach the impurity site. This is shown in Fig. 4, which is series of conductance spectra as a function of distance from the Ni site. The particle-like coherence peak is depleted to provide spectral density for the impurity-state, but the gap magnitude (28 mV) deduced from the hole-like coherence peak location is unperturbed.

In summary, two local states in excellent qualitative agreement with d-wave impurity scattering theory can be identified at each Ni impurity atom. The presence of two impurity-states indicates that the Ni atom possesses a magnetic moment, while the existence and structure of the particle-like to hole-like oscillations and the constant gap magnitude indicate that superconductivity is nowhere disrupted.

We now turn to the implications of these data for proposals[5,6] that differences between the influence of "magnetic" Ni and "non-magnetic" Zn impurity atoms are evidence for a magnetic mechanism in HTSC. Previous experimental studies using Ni and Zn impurities have revealed two apparently contradictory aspects of their effects on HTSC. On one hand, resistivity[21,22], microwave surface resistivity[23], and $T_c$[21,22] measurements show quite similar dependences on Zn and Ni doping densities. On the other hand, probes sensitive to magnetic phenomena such as NMR[24-30] and inelastic neutron scattering (INS)[31], and probes of the superfluid density such as penetration depth[23] and muon-spin-rotation (μSR)[32], show characteristics of Ni- and Zn-doped samples that are dramatically different.

Analysis of the Ni impurity-state energies using the model of Ref. 8 is the first step towards understanding this situation. In that model, potential scattering is represented by an on-site energy U, while magnetic interaction is modeled as having energy $W = J\,\mathbf{S}\cdot\mathbf{s}$. Here $\mathbf{S}$ is the classical spin of the impurity atom, $\mathbf{s}$ is the quasi-particle spin, and $J$ is the exchange energy. The solution for the on-site resonance energy can be written approximately as[8]:

$$\frac{\Omega_{1,2}}{\Delta_0} = \frac{-1}{2N_F(U \pm W)\ln|8N_F(U \pm W)|} \quad (1)$$

Here $\Delta_0$ is the maximum magnitude of the superconducting gap and $N_F$ is the normal density of states per site at the Fermi energy. By substituting the measured values of $\Omega_1$ and $\Omega_2$ and $\Delta_0 = 28$ meV into this equation, we find $N_F U = -0.67$ and $|N_F W| = 0.14$. This represents a surprising new insight because Ni is usually regarded as a source of magnetic scattering but here we identify the dominant effects as due to potential scattering.

The apparent conflicts between results from different probes may now be reconciled. The first set of probes is sensitive to scattering. Calculation of the potential scattering phase shift $\delta_0 = \tan^{-1}(\pi N_F U)$ for Ni gives $\delta_0 = 0.36\pi$ while our previous results showed that Zn is a unitary scatterer ($\delta_0 \approx \pi/2$)[9]. The similarity of these phase shifts imply that phenomena dependent on scattering should be quite similar in Ni- and Zn-doped samples. In fact, using these parameters in an Abrikosov-Gorkov model[1] (and ignoring Ni's magnetic potential) we calculate that $T_C$ would be suppressed only about 20% faster by Zn than by Ni, certainly within the range of experimental observations[21,22].

Among the second set of measurement techniques are those sensitive to superconductivity itself. For example, penetration depth[23] and μSR[32] measurements show that, in the bulk, Zn strongly depletes superfluid density but Ni has a much weaker impact. The STM data now provide a microscopic explanation for these differences - Zn atoms locally destroy superconductivity[9] while Ni atoms do not.

Finally, magnetic probes sensitive to spin fluctuations reveal dramatic changes with Zn-doping[24,26-31] but only weak perturbations with Ni-doping[24,25,29,31]. Explanations for these phenomena have been proposed[5,6,27,28,30] whereby Zn behaves like a "magnetic hole" (a spinless site in an environment of strongly exchange-coupled spins) that dramatically alters NN exchange correlations and disrupts superconductivity, while Ni retains a magnetic moment that barely perturbs the antiferromagnetic exchange correlations which facilitate superconductivity. While the NMR and INS data[24-31] are quite consistent with the magnetic component of such models, their predictions for *local electronic* phenomena at Ni and Zn can only now be tested for the first time. The STM data demonstrate that, despite their magnetic moments, scattering at Ni atoms is dominated by potential interactions. Furthermore, whereas Zn atoms locally destroy superconductivity within a 15 Å radius[9], the magnetic Ni atoms coexist with unweakened superconductivity. All these phenomena are consistent with the above proposals.

The resilience of cuprate-oxide high-$T_C$ superconductivity against what should be the destructive effects of a magnetic impurity atom, and its concomitant vulnerability to destruction by a "magnetic hole," are remarkable. These atomic-scale phenomena are now (through a combination of NMR, μSR, and STM experiments) coming into much clearer focus. They point towards a new approach to studying HTSC in which microscopic theories can be tested against an atomically resolved knowledge of impurity-state phenomena.

We acknowledge H. Alloul, P.W. Anderson, A.V. Balatsky, D. Bonn, M. Flatté, M. Franz, D.-H. Lee, K. Maki, I. Martin, P. Monthoux, A. Mourachkine, D. Pines, D. Rokhsar, S. Sachdev, D.J. Scalapino, and A. Yazdani for helpful conversations and communications. We also thank J.E. Hoffman for her help with data analysis. Support was from the Department of Energy, the Office of Naval Research, the UCDRD Program, Grant-in-Aid for Scientific Research on Priority Area (Japan), a COE Grant from the Ministry of Education, Japan, the Miller Inst. for Basic Research (JCD), and by the IBM Graduate Fellowship Program (KML).

Correspondence should be addressed to jcdavis@physics.berkeley.edu or see http://socrates.berkeley.edu/~davisgrp/.

Figure 1: Maps of the local density of electronic states, at energies E=±9meV revealing the locations of the Ni impurity states

Two 128 Å square differential conductance maps of Ni doped BSCCO at sample-bias (a) + 9 mV, showing the "✚-shaped" regions of high local density of states (LDOS) associated with the Ni atoms and (b) – 9 mV, showing the 45° spatially rotated "✘-shaped" pattern. The differential conductance G(V) = dI/dV measured at sample bias voltage V and tunneling current I is proportional to the LDOS at energy E = eV. In such a LDOS map, states associated with positive sample-bias are referred to as particle-like while those at negative sample-bias are hole-like. There are approximately 8 Ni atom sites in the field of view. Since the areal densities of these phenomena in two different crystals are in good agreement with the two different Ni dopant densities, we conclude that they are associated with Ni atoms. The junction resistance was 1 GΩ at – 100 mV for both maps and all data reported in this paper were acquired at 4.2 K.

Figure 2: Detailed spatial structure of the impurity-state at a single Ni atom.

(a, b) 35 Å square differential conductance maps of a Ni atom at (a) + 9 mV and (b) – 9 mV. (c) A 35 Å square atomic resolution topograph of the BiO surface obtained simultaneously with the maps. (d) Using the information in Figs. (a), (b) and (c) and the fact that the Cu atoms in the CuO$_2$ plane lie directly beneath the Bi atoms in the BiO plane, we draw a schematic showing the relative position of the Ni atom (black solid circle) with respect to the Cu atoms (orange solid circles) in the invisible CuO$_2$ plane. Also depicted is the orientation of the d$_{x^2-y^2}$ order parameter with respect to the **a**- and **b**-axis, and the positions of the first NN (open circle) and second NN (open square) atoms where the spectra shown in Fig. 3 are acquired. The 10 Å square region inside which the spatial integration of the whole impurity-state LDOS was carried out is also shown. The junction resistance is 1 GΩ at –100 mV for all data. Note that the bright central maximum of the "+-shaped" LDOS map always occurs at the site of a Bi surface atom, consistent with the Ni atom being at the Cu site ~5 Å below. We note that the spatial structure in LDOS of an impurity-state, when observed at the BiO plane, may not be identical to that in the $CuO_2$ plane[16]. However, the magnitude of the impurity-state energies should be independent of interlayer matrix element effects.

Figure 3: Differential conductance spectra above the Ni atom and at several nearby locations.

Differential conductance spectra obtained at four positions nearby the Ni atom showing the resonances at $\pm \Omega_1$ and $\pm \Omega_2$ and the superconducting coherence peaks. These curves are shifted from each other by 1nS for clarity. (a) The spectrum above the Ni atom site (solid circles), above the first nearest neighbor Cu atom position (open circles), above the second nearest neighbor Cu atom (open squares), and a typical spectrum 30 Å from an impurity atom (solid triangles). The location of the Ni atom along with the first and second nearest neighbors is shown schematically in Fig. 2(d). The spectra show rapid variations in LDOS magnitude as a function of position. They also reveal the persistence of the coherence peak (shown by the upward pointing arrows) in any given spectrum on the side opposite to the impurity-state resonance peaks. (b) The average of conductance spectra over a 10 Å square area centered on a Ni atom, showing the particle-hole symmetric nature of the spatially integrated LDOS. The junction resistance was 1 GΩ at –200 mV for all spectra.

Figure 4: Evolution of the dI/dV spectra and superconducting energy gap as a function of distance from the Ni atom.

A series of differential conductance spectra, taken as a function of distance away from the center of a Ni impurity-state along the b-axis. The top spectrum is taken above the Ni site. The spectra, each separated by 0.5 Å, show that over a ~20 Å range the magnitude of the superconducting gap $\Delta_0 = 28$ meV, as measured by the distance from the hole-like coherence peak to the Fermi level, is not suppressed in the presence of the Ni impurity. Line cuts approaching the Ni site from other directions lead to the same conclusion. There are variations in $\Delta_0$ throughout the crystal but this study was carried out in a patch with constant $\Delta_0$ of 28 meV. The junction resistance was 1 GΩ at –100 mV for all spectra.

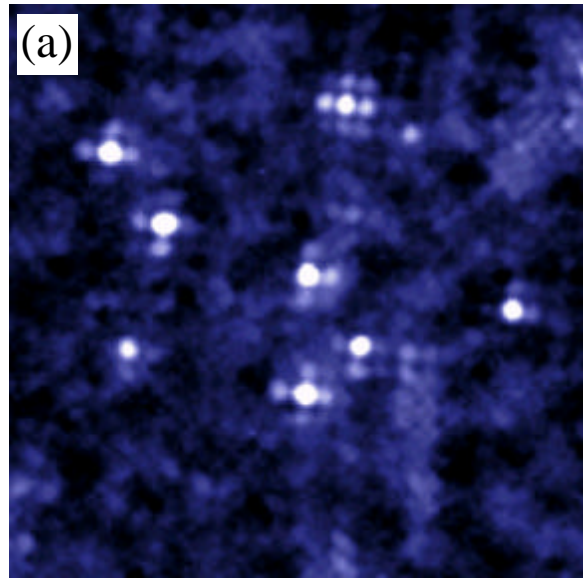
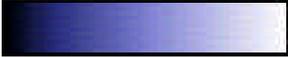
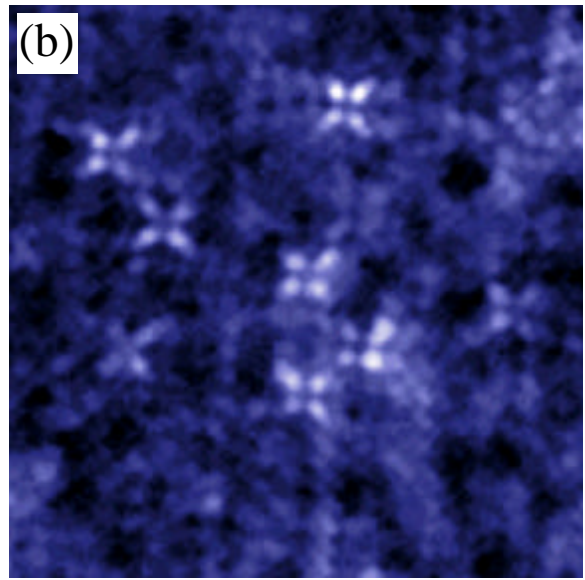

Figure 1
E.W. Hudson *et al.*

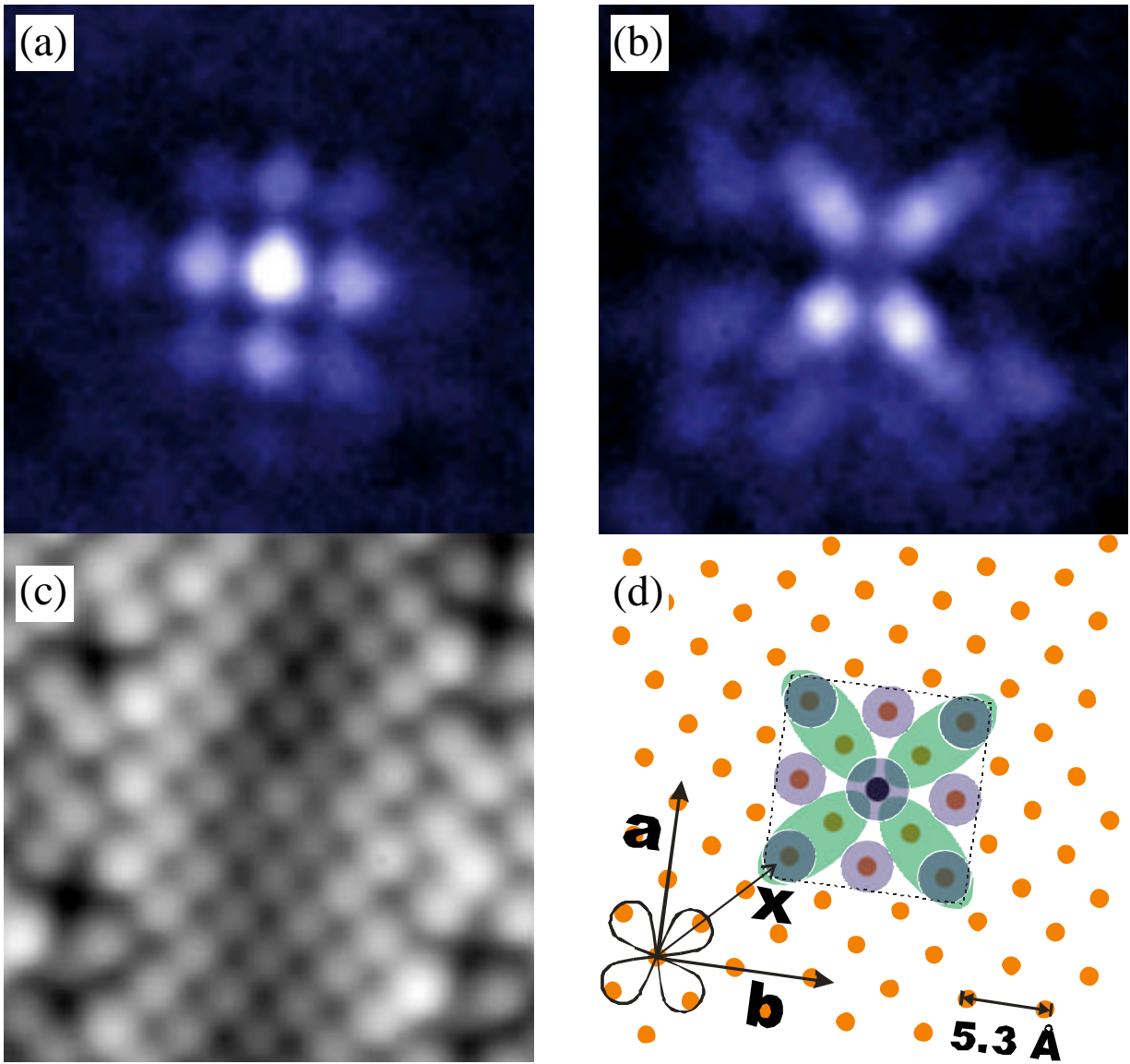

Figure 2
E.W. Hudson *et al.*

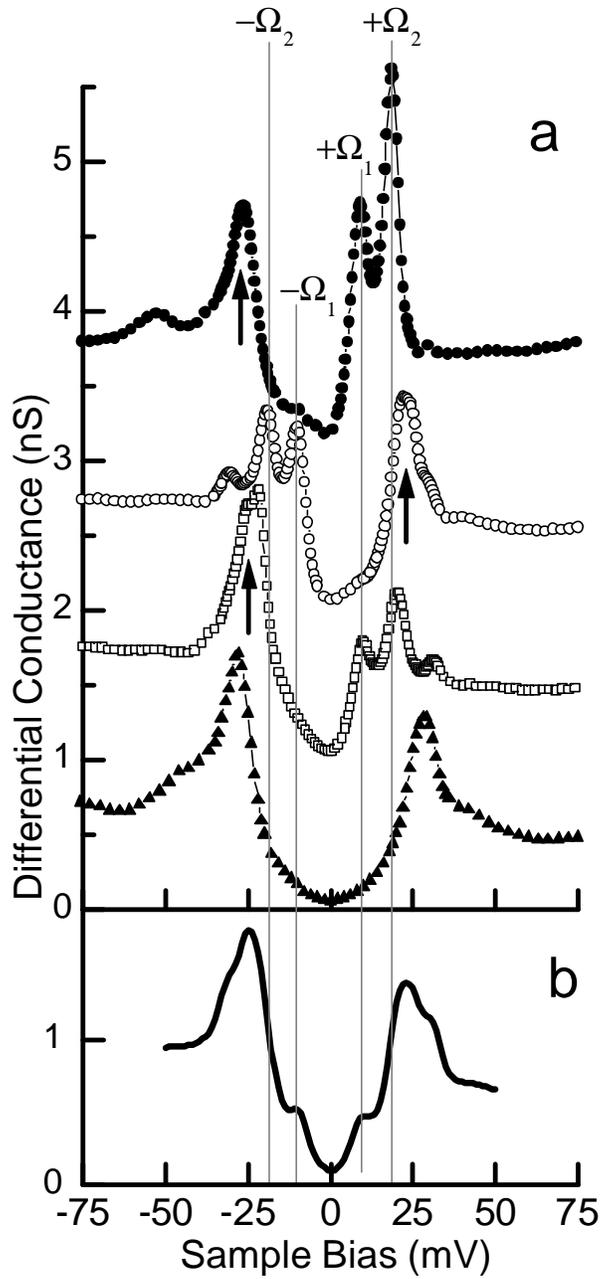

Figure 3
E.W. Hudson *et al.*

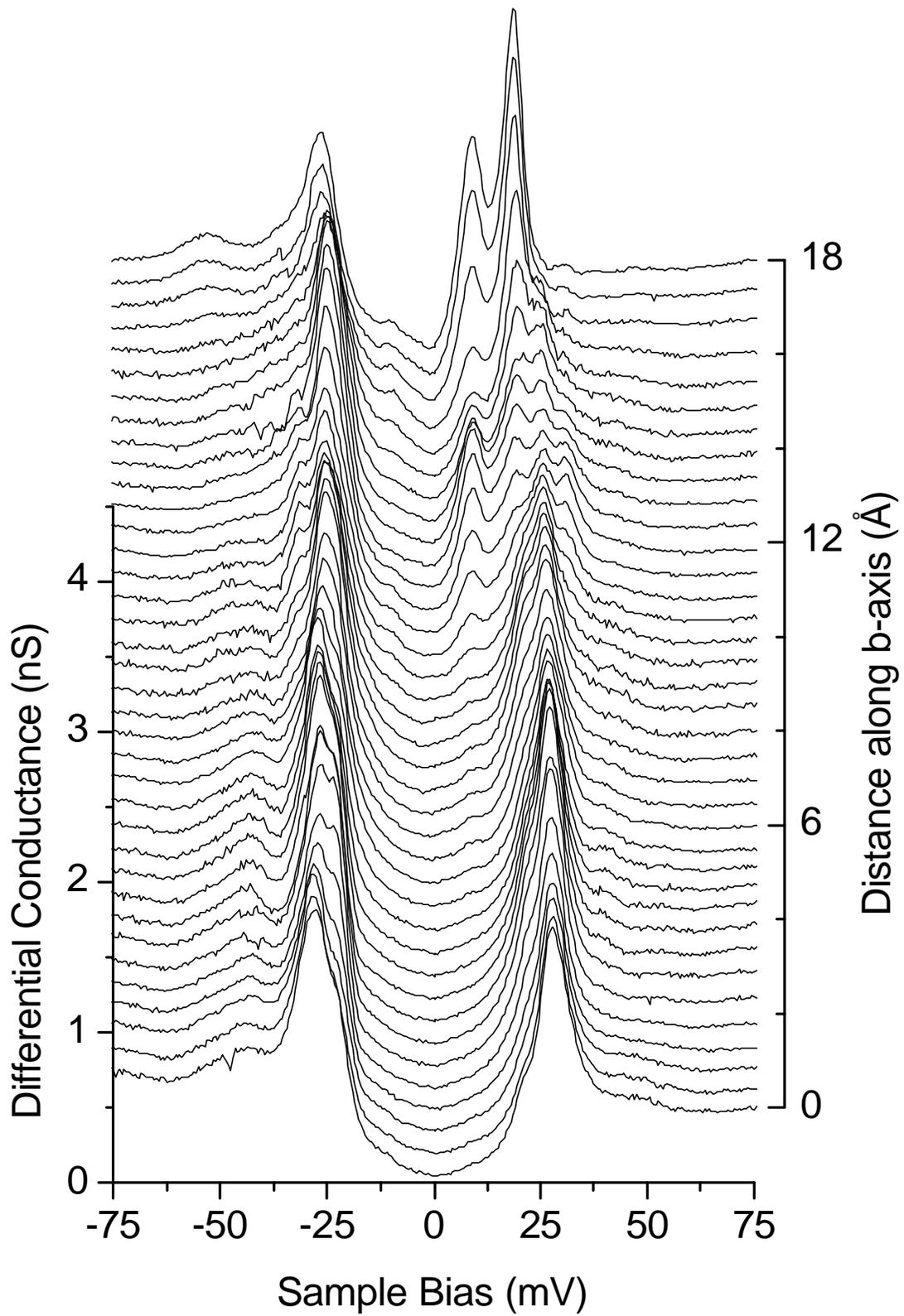

Figure 4
E.W. Hudson *et al.*